\documentclass[aps,prd,twocolumn,superscriptaddress,nofootinbib]{revtex4-2}

\usepackage{amsmath,amssymb,bm}
\usepackage{hyperref}
\usepackage{physics}
\usepackage{booktabs}
\usepackage{siunitx}
\usepackage{graphicx}
\usepackage{adjustbox}
\usepackage{float}
\usepackage{placeins}

\begin{document}

\title{Finite Nuclear Size Corrections on Hyperfine Structure in Muonic Atoms}

\author{Do\u{g}a Ya\c{s}ar}
\affiliation{Max Planck Institute for Nuclear Physics, Saupfercheckweg 1, 69117 Heidelberg, Germany}
\affiliation{Department of Physics, Bo\u{g}azi\c{c}i University, Istanbul, Turkey}

\author{Bastian Sikora}
\affiliation{Max Planck Institute for Nuclear Physics, Saupfercheckweg 1, 69117 Heidelberg, Germany}

\begin{abstract}
Finite nuclear size (FNS) effects on the magnetic-dipole hyperfine splitting in muonic hydrogenlike ions are investigated within a fully relativistic Dirac framework. The FNS contribution is quantified through the correction factor $\delta$, defined by $\Delta E_{\mathrm{ext}} = \Delta E_{\mathrm{point}}(1 - \delta)$, where $\Delta E_{\mathrm{ext}}$ is evaluated using Dirac wavefunctions computed for an extended nuclear charge distribution.

Two nuclear models are considered: a homogeneously charged sphere and a two-parameter Fermi distribution. Bound-state energies and radial wavefunctions are obtained using a numerical iterative solver, while a semi-analytic matching scheme provides reference values and initial seeds. We present a systematic dataset of $\delta$ values for the $1s$, $2s$, and $2p_{1/2}$ states over a wide range of nuclear charge numbers $Z$. Nuclear-model dependence is quantified, including uncertainties induced by the nuclear radius in the uniform-sphere model.

The results show that $\delta$ increases monotonically with $Z$ and exhibits clear state dependence, with reduced magnitude for the $2p_{1/2}$ state relative to $s$ states. A pronounced sensitivity to the nuclear charge distribution is observed, highlighting the importance of realistic nuclear modeling in precision hyperfine studies of muonic atoms.
\end{abstract}

\maketitle

\section{Introduction}

A muonic atom is an effectively two-body system consisting of a negatively charged muon bound to a nucleus. The muon is a spin-$1/2$ lepton with the same quantum numbers as the electron but with a significantly larger mass, $m_{\mu} \approx 207\, m_e$. Owing to the inverse scaling between mass and Bohr radius, the increased mass of the muon leads to a substantial contraction of the atomic orbit. As a consequence, muonic wavefunctions exhibit a much stronger overlap with the nuclear region than electronic ones. This enhanced sensitivity makes muonic atoms particularly suitable systems for probing nuclear structure and extracting nuclear parameters~\cite{Wu1969,Schaller1980,Ruetschi1984,Piller1990,Pohl2010,Michel2017,Patoary2018,Michel2019,Antognini2020,Paul2021,Okumura2021,Valuev2022,Oreshkina2022,Saito2022,Yerokhin2023,Sun2025,Vandeleur2025,Quint2026}.

Atomic observables in such systems become directly sensitive to the nuclear charge distribution through the behavior of the bound-state wavefunction at small radii. One of the most important of these observables is the hyperfine structure (HFS), which arises from the interaction between the magnetic moment of the nucleus and the magnetic field generated by the bound lepton~\cite{Antognini2013,Antognini2020,Pohl2022,Pohl2023}. While hyperfine splitting constitutes a relatively small correction in ordinary electronic atoms, it is significantly enhanced in muonic systems due to the increased wavefunction density near the nucleus.

In the present work, the finite nuclear size (FNS) contribution to the magnetic dipole hyperfine interaction is investigated for the $1s$, $2s$, and $2p_{1/2}$ states of muonic ions. The FNS contribution is denoted by $\delta$ and is defined through the relation
\begin{equation}
\Delta E_{\mathrm{ext}} = \Delta E_{\mathrm{point}} (1 - \delta),
\end{equation}
where $\Delta E_{\mathrm{point}}$ and $\Delta E_{\mathrm{ext}}$ correspond to the hyperfine splitting obtained for a point-like and an extended nuclear charge distribution, respectively.

Two different charge distribution models are considered: the homogeneously charged sphere and the two-parameter Fermi distribution. For the homogeneously charged sphere model, the uncertainty in $\delta$ originating from the uncertainty in the root-mean-square nuclear radius $R_{\mathrm{rms}}$ is evaluated to assess the sensitivity of the hyperfine correction to nuclear size parameters. In contrast to previous studies, the present work provides a systematic and numerically stable evaluation of FNS corrections across a wide range of $Z$ within a unified relativistic framework.

The magnetic dipole hyperfine interaction is treated within the fully relativistic Dirac framework. The radial matrix element entering the magnetic dipole interaction,
\[
\int_0^\infty f_{n\kappa}(r)\,g_{n\kappa}(r)\,dr ,
\]
is evaluated using the radial Dirac wavefunctions $g_{n\kappa}(r)$ and $f_{n\kappa}(r)$ obtained for the finite nuclear charge distribution. The resulting energy shift is matched to the standard point-nucleus hyperfine expression, from which the finite nuclear size correction $\delta$ is determined. Only the modification arising from the finite nuclear charge distribution is considered; effects associated with the spatial distribution of the nuclear magnetization (Bohr--Weisskopf correction) are not included.

Relativistic units $(\hbar = c = 1)$ and Heaviside--Lorentz charge units $(\alpha = e^2/4\pi)$ are used throughout the paper, where $\alpha$ denotes the fine-structure constant and $e<0$.

In this context, eigenenergies, radial wavefunctions, and the corresponding finite nuclear size contribution to hyperfine structure are systematically analyzed and presented. The paper is organized as follows:
Section~\ref{sec:theoretical_framework} introduces the theoretical framework and relevant analytical structures;
Section~\ref{sec:numerical_methods} describes the numerical methods developed for the solution of the problem;
numerical results are presented in Section~\ref{sec:data} and analyzed in Section~\ref{sec:analysis};
finally, conclusions are given in Section~\ref{sec:conclusion}.
\section{Theoretical Framework}
\label{sec:theoretical_framework}
\subsection{Dirac Equation in a Central Potential}
\label{sec:dirac_equation}

The relativistic bound states of a spin-$1/2$ fermion in an external field are described by the stationary Dirac equation
\begin{equation}
\left[ \boldsymbol{\alpha}\cdot \mathbf{p} + \beta m_0 + V(r) \right]\psi(\mathbf{r}) = E \psi(\mathbf{r}),
\end{equation}
where $m_0$ denotes the rest mass of the fermion, $\boldsymbol{\alpha}$ and $\beta$ are the Dirac matrices, and $V(r)$ is a central potential. The eigenvalues $E$ correspond to the bound-state energies, while the eigenfunctions $\psi(\mathbf{r})$ represent the associated wavefunctions.

For a central potential $V(\mathbf{r}) = V(r)$, the wavefunction can be separated into radial and angular parts. In standard representation, it can be written as
\begin{equation}
\psi_{n\kappa m_j}(\mathbf{r}) =
\frac{1}{r}
\begin{pmatrix}
G_{n\kappa}(r)\,\Omega_{\kappa m_j}(\theta,\phi) \\
i F_{n\kappa}(r)\,\Omega_{-\kappa m_j}(\theta,\phi)
\end{pmatrix},
\label{eq:dirac_spinor}
\end{equation}
where $G_{n\kappa}(r)$ and $F_{n\kappa}(r)$ denote the large and small radial components, respectively. The functions $\Omega_{\kappa m_j}(\theta,\phi)$ are the spinor spherical harmonics describing the angular dependence.

The relativistic quantum number $\kappa$ is defined as
\begin{equation}
\kappa =
\begin{cases}
-(j + \tfrac{1}{2}), & \text{for } j = l + \tfrac{1}{2}, \\
+(j + \tfrac{1}{2}), & \text{for } j = l - \tfrac{1}{2},
\end{cases}
\end{equation}
where $l$ is the orbital angular momentum, $s=1/2$ is the spin of the fermion, and $j = l \pm 1/2$ denotes the total angular momentum. The magnetic quantum number $m_j$ corresponds to the projection of the total angular momentum along the quantization axis.

Substituting the above ansatz into the Dirac equation yields the coupled radial equations~\cite{GreinerRQM2000}
\begin{align}
\frac{dG}{dr} + \frac{\kappa}{r} G(r)
- \bigl[m_0 - V(r)\bigr] F(r) &= E F(r), \\
-\frac{dF}{dr} + \frac{\kappa}{r} F(r)
+ \bigl[m_0 + V(r)\bigr] G(r) &= E G(r).
\label{eq:coupled_dif_eq}
\end{align}
The explicit form of the potential $V(r)$ encodes the nuclear charge distribution and therefore determines the short-distance behavior of the radial wavefunctions. In the following subsection, different nuclear charge models are introduced.
\subsection{Nuclear Charge Distributions}
\label{sec:charge_distribution}

The nuclear charge distribution determines the form of the central potential $V(r)$ entering the Dirac equation. In the point-like limit, the nucleus generates the Coulomb potential
\begin{equation}
V_C(r) = -\frac{Z\alpha}{r},
\end{equation}
where $Z$ is the nuclear charge number and $\alpha$ is the fine-structure constant.

\subsubsection*{Homogeneously Charged Sphere}

As a simple extended-nucleus model, the nucleus can be approximated by a homogeneously charged sphere with radius $R_0$. The corresponding charge density is
\begin{equation}
\rho(r) = \frac{-3Ze}{4\pi R_0^3}\,\theta(R_0-r),
\end{equation}
where $\theta$ denotes the Heaviside step function. For a spherically symmetric charge distribution, the root-mean-square nuclear radius $R_{\mathrm{rms}}$ and the corresponding effective radius $R_0$ are given by
\begin{equation}
R_{\mathrm{rms}}=\left(\frac{\int_0^\infty r^4 \rho(r)\,dr}{\int_0^\infty r^2 \rho(r)\,dr}\right)^{1/2},
\qquad
R_0=\sqrt{\frac{5}{3}}\,R_{\mathrm{rms}}.
\label{eq:r0}
\end{equation}
Then the electrostatic potential is
\begin{equation}
V(r)=
\begin{cases}
-\dfrac{Z\alpha}{2R_0}\left(3-\dfrac{r^2}{R_0^2}\right), & r\le R_0, \\[6pt]
-\dfrac{Z\alpha}{r}, & r>R_0.
\end{cases}
\label{eq:dirac_potential}
\end{equation}

\subsubsection*{Two-Parameter Fermi Distribution}

A more realistic description of the nuclear charge density is provided by the two-parameter Fermi distribution,
\begin{equation}
\rho(r) = \frac{N}{1+\exp\!\left(\dfrac{r-c}{a}\right)},
\end{equation}
where $c$ is the half-density radius and $a$ is the diffuseness parameter. In this work, the surface thickness parameter $t$ is defined as
\begin{equation}
t = 4\ln 3\, a,
\end{equation}
and the normalization constant $N$ is fixed by the condition
\begin{equation}
\int \rho(r)\, d^3r = Ze .
\end{equation}
Evaluating this condition for the two-parameter Fermi distribution
yields the explicit expression
\begin{equation}
N = \frac{3}{4\pi c^3}
\left(1+\frac{\pi^2 a^2}{c^2}\right)^{-1},
\end{equation}
which follows from the analytic treatment of the Fermi model
(see, e.g., Ref.~\cite{Beier2000}).
The parameter $c$ is related to the RMS radius through
\begin{equation}
c^2=\frac{5}{3}R_{\mathrm{rms}}^2-\frac{7}{3}\pi^2 a^2.
\label{eq:c}
\end{equation}
For the numerical implementation, the surface thickness is taken as $t=2.3~\mathrm{fm}$, following Ref.~\cite{Beier2000}. For nuclei with $R_{\mathrm{rms}}<2~\mathrm{fm}$, however, this choice would lead to a negative value of the Fermi parameter $c$ in Eq.~\eqref{eq:c}. Therefore, for these nuclei, a reduced value $t=0.9~\mathrm{fm}$ is adopted.

Following the closed-form expressions given in Ref.~\cite{Beier2000}, the electrostatic potential corresponding to the two-parameter Fermi distribution can be written in analytic form in terms of auxiliary series $S_k(x)$. Defining
\begin{align}
S_k(x) &= \sum_{m=1}^{\infty}\frac{(-1)^m e^{mx}}{m^k}, \\
K &= 1+\frac{\pi^2 a^2}{c^2}-6\left(\frac{a}{c}\right)^3 S_3\!\left(-\frac{c}{a}\right),
\end{align}
the electrostatic potential for the two-parameter Fermi distribution can be given as
\begin{align}
V_{\mathrm{F}}(r) &=
-\frac{Z\alpha}{K r}\, \mathcal{F}_{\mathrm{in}}(r),
\qquad r \le c,
\end{align}
where
\begin{align}
\mathcal{F}_{\mathrm{in}}(r) &=
-6\left(\frac{a}{c}\right)^3
S_3\!\left(-\frac{c}{a}\right)
+6\left(\frac{a}{c}\right)^3
S_3\!\left(\frac{r-c}{a}\right)
+\frac{r}{c}
\nonumber\\
&\quad \times
\left[
\frac{3}{2}
+\frac{\pi^2}{2}\left(\frac{a}{c}\right)^2
-3\left(\frac{a}{c}\right)^2
S_2\!\left(\frac{r-c}{a}\right)
-\frac{1}{2}\left(\frac{r}{c}\right)^2
\right].
\end{align}

For $r>c$, one obtains
\begin{align}
V_{\mathrm{F}}(r) &=
-\frac{Z\alpha}{r}
-\frac{3Z\alpha}{K r}
\left(\frac{a}{c}\right)^2
\mathcal{F}_{\mathrm{out}}(r),
\qquad r > c,
\end{align}
where
\begin{align}
\mathcal{F}_{\mathrm{out}}(r) &=
\frac{r}{c}
S_2\!\left(\frac{c-r}{a}\right)
+2\left(\frac{a}{c}\right)
S_3\!\left(\frac{c-r}{a}\right).
\end{align}

\begin{figure}[h]
\centering
\includegraphics[width=\columnwidth]{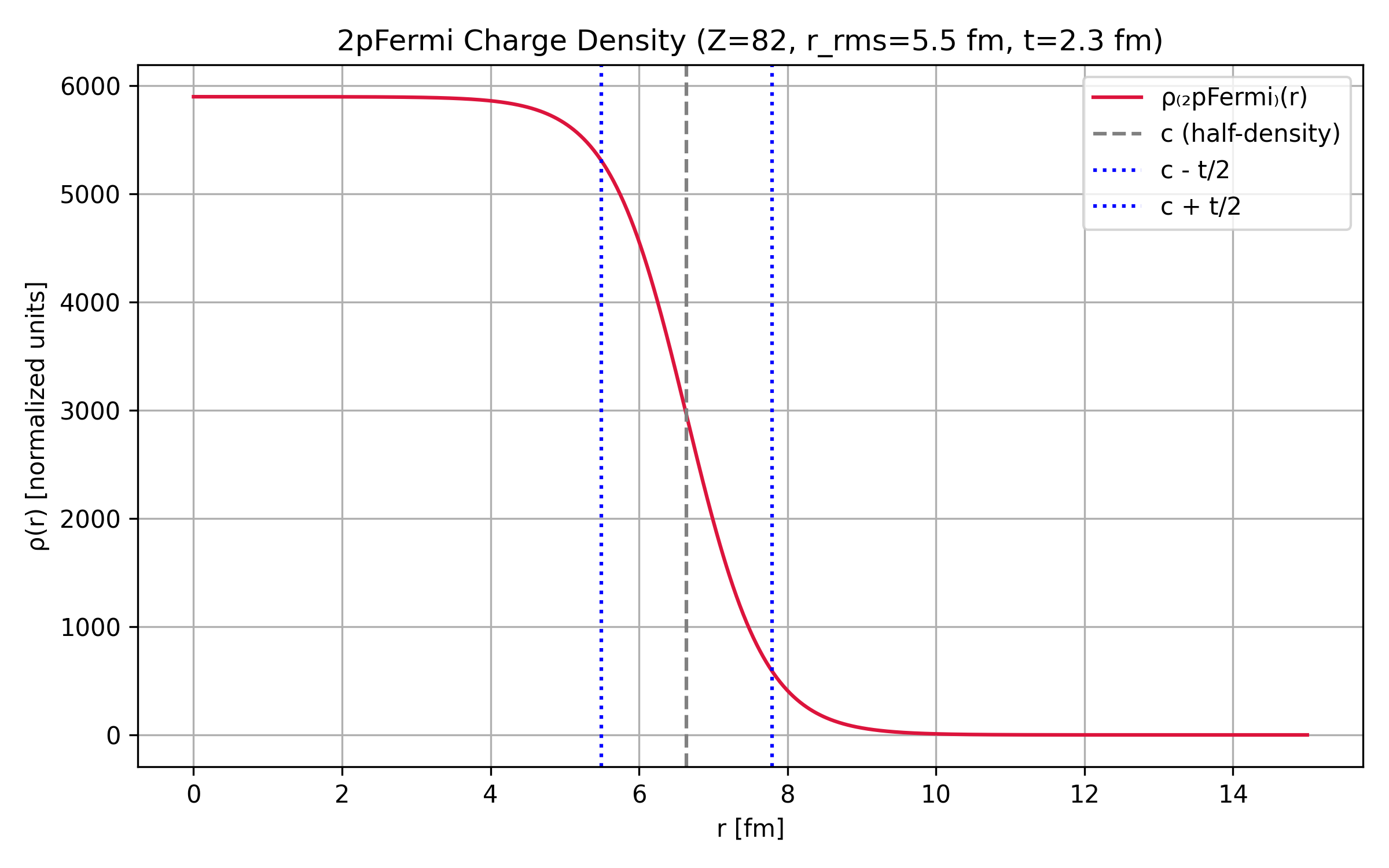}
\caption{Radial dependence of the two-parameter Fermi charge distribution for representative nuclear parameters.}
\label{fig:rho_2pfermi}
\end{figure}

Figure~\ref{fig:rho_2pfermi} illustrates the radial dependence of the two-parameter Fermi charge distribution for representative nuclear parameters.

\subsection{Radial Solutions for Extended Nuclei}
\label{sec:radial_solutions}

When the homogeneously charged sphere model is employed, the potential is given by Eq.(\ref{eq:dirac_potential}). 
The radial Dirac equation can then be solved separately in the inner region ($r \le R_0$) and outer region ($r > R_0$)~\cite{Patoary2018}.

\subsubsection*{Region I ($r \le R_0$)}

Inside the nucleus, the regular solution at the origin can be written as a power-series expansion,

\begin{equation}
\begin{pmatrix}
G(r) \\
F(r)
\end{pmatrix}
=
N_1\, r^{|\kappa|}
\sum_{i=0}^{\infty}
\left[
a_i \pm (-1)^{i+1}\frac{\kappa}{|\kappa|} a_i
\right]
r^i,
\label{eq:region1_series}
\end{equation}

Unless stated otherwise, the upper (lower) sign refers to the $G$ ($F$) component,
where $N_1$ is a normalization constant and the coefficients $a_i$ are determined recursively.

The recursion relation is

\begin{equation}
a_i =
\frac{
a_{i-1}
\left[
E + \frac{3Z\alpha}{2R_0}
- m_0 (-1)^i \frac{\kappa}{|\kappa|}
\right]
-
\frac{Z\alpha}{2R_0^{3}}\, a_{i-3}
}
{
\kappa + (-1)^{i+1}\frac{\kappa}{|\kappa|}(i + |\kappa|)
},
\label{eq:ai_recursion}
\end{equation}

with initial conditions
\[
a_i = 0 \quad (i < 0), 
\qquad
a_0 = 1.
\]

\subsubsection*{Region II ($r > R_0$)}

Outside the nucleus, the potential reduces to the Coulomb form and the radial solutions can be expressed in terms of Whittaker functions.

We introduce the dimensionless variable
\begin{equation}
\rho = 2 r \sqrt{m_0^2 - E^2} \equiv 2 \lambda r,
\label{eq:rho_lambda}
\end{equation}
with $\lambda = \sqrt{m_0^2 - E^2}$.

The bound-state solution regular at infinity is given by
\begin{equation}
\begin{aligned}
\begin{pmatrix}
G(\rho) \\
F(\rho)
\end{pmatrix}
&=
\frac{N_2}{\kappa + \dfrac{m_0 Z\alpha}{\lambda}}
\rho^{-1/2}
\begin{pmatrix}
\sqrt{m_0+E} \\
\sqrt{m_0-E}
\end{pmatrix}
\\
&\quad \times
\left[
\left(\kappa + \frac{m_0 Z\alpha}{\lambda}\right)
W_{q,\gamma}(\rho)
\pm
W_{q+1,\gamma}(\rho)
\right],
\end{aligned}
\label{eq:region2_solution}
\end{equation}
where
\[
q = \frac{Z\alpha E}{\lambda} - \frac{1}{2},
\qquad
\gamma = \sqrt{\kappa^2 - (Z\alpha)^2}.
\]
Here $W_{q,\gamma}(\rho)$ denotes the Whittaker function of the second kind.

\subsubsection*{Matching Conditions and Energy Equation}

The physical wavefunctions must satisfy normalization and continuity conditions,

\begin{equation}
\int_0^{\infty} \left(G^2 + F^2\right) dr = 1,
\end{equation}

\begin{equation}
G(R_0^-)=G(R_0^+),
\qquad
F(R_0^-)=F(R_0^+).
\end{equation}

By inserting the inner (Region I) and outer (Region II) solutions at $r=R_0$, one obtains a transcendental equation for the energy~\cite{Patoary2018},

\begin{equation}
\begin{split}
\frac{
A_1 \left(\kappa + \dfrac{m_0 Z\alpha}{\lambda}\right)
W_{q,\gamma}(2\lambda R_0)
+
A_2 W_{q+1,\gamma}(2\lambda R_0)
}{
A_2 \left(\kappa + \dfrac{m_0 Z\alpha}{\lambda}\right)
W_{q,\gamma}(2\lambda R_0)
+
A_1 W_{q+1,\gamma}(2\lambda R_0)
}
\\[6pt]
=
\frac{
\sum_{i=0}^{\infty} a_i R_0^i
}{
\sum_{i=0}^{\infty}
(-1)^{i+1}\frac{\kappa}{|\kappa|}
a_i R_0^i
}.
\end{split}
\label{eq:energy_condition}
\end{equation}

where

\begin{equation}
A_{1,2} = \sqrt{m_0 + E} \pm \sqrt{m_0 - E}.
\end{equation}

The transcendental Eq.(\ref{eq:energy_condition}) determines the bound-state energies and provides the reference values used in the numerical analysis.

\subsection{Magnetic Dipole Hyperfine Interaction}

The hyperfine structure (HFS) arises from the interaction between the magnetic moment of the nucleus and the magnetic field generated by the bound muon. 
The interaction Hamiltonian can be written as~\cite{Beier2000}

\begin{equation}
H_{\mathrm{HFS}} = - \boldsymbol{\mu}_I \cdot \mathbf{B}_J 
= - \boldsymbol{\mu}_J \cdot \mathbf{B}_I ,
\end{equation}
where $\boldsymbol{\mu}_I$ and $\boldsymbol{\mu}_J$ denote the magnetic moment operators of the nucleus and the bound muon, respectively, and $\mathbf{B}$ is the magnetic field.

The magnetic vector potential generated by a nuclear magnetic dipole reads

\begin{equation}
\mathbf{A}_{\mathrm{HFS}}(\mathbf{r}) 
= \frac{\boldsymbol{\mu}_I \times \mathbf{r}}{4\pi r^3}.
\end{equation}

The magnetic moments are given by

\begin{equation}
\boldsymbol{\mu}_J = - g_J \frac{e}{2m_0}\mathbf{J},
\qquad
\boldsymbol{\mu}_I = g_I \frac{e}{2m_p}\mathbf{I}.
\end{equation}
where $m_p$ is the proton mass, $g_J$ and $g_I$ are the $g$-factors of the muon and nucleus, respectively, and $\mathbf{J}$ and $\mathbf{I}$ are their angular momentum operators.

The energy shift due to the magnetic dipole hyperfine interaction is obtained from the expectation value

\begin{equation}
\Delta E_{\mathrm{HFS}}
=
\langle a_{n,F,m_F} | e\, \boldsymbol{\alpha}\!\cdot\!\mathbf{A}_{\mathrm{HFS}} | a_{n,F,m_F} \rangle .
\end{equation}

Performing the angular integration yields

\begin{equation}
\begin{aligned}
\Delta E_{\mathrm{HFS}}^{\mathrm{(ext)}}
&=
\frac{\alpha\,\hbar}{2m_p}\,
g_I\,
\frac{4\kappa}{4\kappa^2-1}
\left[
F(F+1) - I(I+1)
\right.
\\
&\qquad\left.
-\, j(j+1)
\right]
\times
\int_0^\infty
f_{n\kappa}(r)\, g_{n\kappa}(r)\, dr ,
\end{aligned}
\label{eq:hfs_general}
\end{equation}
where $f_{n\kappa}(r)$ and $g_{n\kappa}(r)$ denote the small and large radial components of the Dirac wavefunction, respectively.

\noindent
In our notation (see Eq.(\ref{eq:dirac_spinor})), the radial functions are related through
\begin{equation}
g_{n\kappa}(r)=\frac{G_{n\kappa}(r)}{r},
\qquad
f_{n\kappa}(r)=\frac{F_{n\kappa}(r)}{r}.
\label{eq:reduced_components}
\end{equation}

\subsection{Finite Nuclear Size Correction to Hyperfine Splitting}
\label{sec:hpf}

For a point-like nucleus, the magnetic dipole hyperfine splitting admits a closed analytic expression. It can be written as~\cite{Beier2000}

\begin{equation}
\begin{aligned}
\Delta E_{\mathrm{HFS}}^{\mathrm{(point)}}
&=
\alpha g_I
\frac{m_0}{m_p}
\frac{
F(F+1)-I(I+1)-j(j+1)
}{
2j(j+1)
}
\\[6pt]
&\quad \times
m_0 c^2
\frac{(Z\alpha)^3}{n^3(2l+1)}
A(Z\alpha).
\end{aligned}
\label{eq:hfs_point}
\end{equation}
where the relativistic factor $A(Z\alpha)$ is given by
\begin{equation}
A(Z\alpha)
=
n^3(2l+1)
\frac{
\kappa \left[ 2\kappa(\gamma+n_r)-N' \right]
}{
N'^4 \gamma (4\gamma^2-1)
}.
\end{equation}

Here the auxiliary quantities are defined as
\begin{equation}
\begin{aligned}
N' &= \sqrt{n_r^2 + 2 n_r \gamma + \kappa^2}, \\
n_r &= n - |\kappa|,
\end{aligned}
\end{equation}
where $\gamma$ is defined in Eq.~\eqref{eq:region2_solution}.

For an extended nucleus, the hyperfine splitting is modified through the change in the radial integral appearing in Eq.~\eqref{eq:hfs_general}. 
As stated in the Introduction, we restrict attention to the finite charge-distribution effect.

We recall the definition of the finite nuclear size correction $\delta$,
\begin{equation}
\Delta E_{\mathrm{HFS}}^{\mathrm{(ext)}}
=
\Delta E_{\mathrm{HFS}}^{\mathrm{(point)}}
(1-\delta),
\label{eq:delta_definition}
\end{equation}
which quantifies the relative modification of the hyperfine splitting induced by the finite nuclear charge distribution.

\section{Numerical Methods}
\label{sec:numerical_methods}

\subsection{Problem setup and units}
\label{sec:setup_units}

All numerical calculations are performed in relativistic units, $\hbar = c = 1$, such that energies are expressed in MeV and lengths in $\mathrm{MeV}^{-1}$. 
The fine-structure constant and lepton masses are taken from the CODATA recommended values~\cite{ref3}, namely
$m_e = 0.51099895069~\mathrm{MeV}$ and 
$m_\mu = 105.6583755~\mathrm{MeV}$.

Experimental nuclear root-mean-square (rms) charge radii $R_{\mathrm{rms}}$ for $Z \ge 2$ are adopted from the evaluated data tables of Angeli and Marinova~\cite{ref2}, while the proton charge radius ($Z=1$) is taken from the CODATA compilation~\cite{ref3}. 
All radii given in femtometers are converted to relativistic units using $\hbar c = 197.3269804~\mathrm{MeV\,fm}$.

The sharp radius $R_0$ and the relativistic Coulomb parameter $\gamma$ are defined by Eqs.~(\ref{eq:r0}) and (\ref{eq:region2_solution}), respectively, and are used consistently in the numerical implementation.
\subsection{Radial grid construction}
\label{sec:radial_grid}

The coupled radial Dirac equations are solved on a composite grid that resolves the rapidly varying wavefunctions in the nuclear and near-nuclear region while keeping the computational cost moderate at large radii.
We split the radial domain into an \emph{inner} interval $[r_{\min}, r_{\mathrm{lim}}]$ and an \emph{outer} interval $[r_{\mathrm{lim}}, r_{\max}]$, where the matching radius is chosen as
\begin{equation}
r_{\mathrm{lim}} \equiv R_0 ,
\end{equation}
with $R_0$ being the effective nuclear radius of the chosen charge model.
The lower bound is set to a small positive value to avoid the singular point $r=0$ in the numerical propagation,
\begin{equation}
r_{\min} = \frac{10^{-20}}{m},
\end{equation}
where $m$ is the bound-lepton mass in relativistic units.

To ensure that the bound-state tail is sufficiently captured for all $Z$ and $(n,\kappa)$ considered, we choose the outer boundary $r_{\max}$ via a $Z$-dependent estimate (see~\cite{Yerokhin2003}),
\begin{equation}
r_{\max} = 
\frac{2\,\gamma(Z)\,80}{m\,\alpha\,Z\,\gamma(80)},
\label{eq:rmax_choice}
\end{equation}
where the relativistic parameter $\gamma$ is defined in Eq.~(\ref{eq:region2_solution}).
This choice scales inversely with the lepton mass and increases for lighter systems, thereby providing a conservative radial extent across the full range of nuclear charges.

The inner and outer subgrids are constructed using a monotonic coordinate mapping based on a hyperbolic tangent, which clusters points near the matching radius $r_{\mathrm{lim}}$ from both sides.
For the inner grid, points are concentrated towards the right endpoint $r_{\mathrm{lim}}$ according to
\begin{equation}
\begin{aligned}
r_i^{\mathrm{(in)}} &= r_{\min} + \eta_i^{\mathrm{(in)}}\,(r_{\mathrm{lim}}-r_{\min}), \\
\eta_i^{\mathrm{(in)}} &= \frac{\tanh(\beta_{\mathrm{in}}\,\xi_i)}{\tanh(\beta_{\mathrm{in}})}, \\
\xi_i &= \frac{i}{N_{\mathrm{in}}-1},
\end{aligned}
\label{eq:grid_inside}
\end{equation}
with $i=0,\dots,N_{\mathrm{in}}-1$. 
For the outer grid, points are concentrated towards the left endpoint $r_{\mathrm{lim}}$ using
\begin{equation}
\begin{aligned}
r_j^{\mathrm{(out)}} &= r_{\max} + \eta_j^{\mathrm{(out)}}\,(r_{\mathrm{lim}}-r_{\max}), \\
\eta_j^{\mathrm{(out)}} &= \frac{\tanh\!\big(\beta_{\mathrm{out}}\,(1-\zeta_j)\big)}{\tanh(\beta_{\mathrm{out}})}, \\
\zeta_j &= \frac{j}{N_{\mathrm{out}}-1},
\end{aligned}
\label{eq:grid_outside}
\end{equation}
with $j=0,\dots,N_{\mathrm{out}}-1$.
This construction yields a smooth increase of $r$ across each subinterval and provides enhanced resolution around $r_{\mathrm{lim}}$, where matching conditions and residuals are evaluated in the iterative solver.

We use $N_{\mathrm{in}}=1000$ points in the inner region and $N_{\mathrm{out}}=8000$ points in the outer region (total of 9000 points), with $\beta_{\mathrm{in}}=3.0$ and $\beta_{\mathrm{out}}=5.0$.
For diagnostic purposes, we also implemented a uniform grid on $[r_{\min},r_{\max}]$; however, the non-uniform grid was found to be substantially more efficient for resolving the matching region near $r_{\mathrm{lim}}$.

\subsection{Semi-analytic eigenvalue solver}
\label{sec:semi_analytic_solver}

The bound-state energies are obtained by solving the transcendental matching equation
given in Eq.~(\ref{eq:energy_condition}). 
For numerical purposes, this equation is rewritten in residual form
\begin{equation}
\mathcal{R}(E) = \mathrm{LHS}(E) - \mathrm{RHS}(E),
\label{eq:residual_definition}
\end{equation}
where LHS and RHS denote the left- and right-hand sides of Eq.~(\ref{eq:energy_condition}).
The eigenenergy corresponds to the root
\begin{equation}
\mathcal{R}(E)=0.
\end{equation}

\paragraph{Radial ingredients of the residual.}

The residual evaluation requires both radial solutions.

(i) The inner solution (Region I) is constructed via the Frobenius (power-series) expansion of Eq.~(\ref{eq:region1_series}), with coefficients determined from the
recursion relation Eq.~(\ref{eq:ai_recursion}).
The series is truncated dynamically once the relative contribution of successive
terms falls below a predefined threshold.
Additional safeguards are implemented to avoid denominator singularities
and uncontrolled coefficient growth.

(ii) The outer solution (Region II) is expressed in terms of Whittaker functions,
Eq.~(\ref{eq:region2_solution}).
In particular, the residual requires the evaluation of
$W_{q,\gamma}(\rho)$ and $W_{q+1,\gamma}(\rho)$
at the matching point $r=R_0$.
Since these functions are evaluated at $\rho_0=2\lambda R_0$, direct numerical evaluation can become unstable for small $\rho_0$, leading to loss of significance and requiring additional stabilization.

To ensure stable residual evaluation, a hybrid computation scheme was implemented.
For sufficiently small $\rho_0$, the leading asymptotic form of the Whittaker
function was used \cite{DLMF_13_14},
\begin{equation}
W_{q,\gamma}(\rho_0)
\sim
\frac{\Gamma(2\gamma)}
{\Gamma\!\left(\gamma-q+\tfrac12\right)}
\, \rho_0^{-\gamma+\tfrac12},
\qquad \rho_0 \to 0,
\end{equation}
which follows from the small-argument expansion of $W_{q,\gamma}$.
This representation stabilizes the evaluation.

For intermediate values of $\rho_0$, the Whittaker function was evaluated
via its confluent hypergeometric representation \cite{DLMF_13_14},
\begin{equation}
W_{q,\gamma}(\rho_0)=
\rho_0^{\gamma+\tfrac12}e^{-\rho_0/2}
U\!\left(\gamma-q+\tfrac12,\;2\gamma+1,\;\rho_0\right),
\end{equation}
where $U(a,b,z)$ denotes the Tricomi function.

For larger $\rho_0$, the direct high-precision implementation was employed directly.
This multi-stage strategy guarantees stable values of the Whittaker functions
at the matching point and therefore stable evaluation of the residual
$\mathcal{R}(E)$.

For diagnostic purposes only, the Whittaker $M$ function was additionally
evaluated to cross-check the numerical behavior of the special-function backend;
it is not used in the eigenvalue condition itself.

\paragraph{Root-finding strategy.}

In most cases, the energy was determined numerically.

As an initial refinement step, a first-order Taylor expansion around the
point-nucleus Dirac energy $E_0$ (see, e.g., Greiner~\cite{GreinerRQM2000}, Ch.9.7) was applied,
\begin{equation}
E_{\mathrm{init}} =
E_0 -
\frac{\mathcal{R}(E_0)}
{\left.\dfrac{d\mathcal{R}}{dE}\right|_{E_0}},
\label{eq:taylor_step}
\end{equation}
where the derivative was computed using a stabilized central-difference
scheme.
This provides an improved starting value for the iterative search.

The Newton method was then used as the primary root-finding algorithm.
For heavy systems where convergence deteriorated,
a Pegasus fallback scheme was implemented to ensure robustness.

\paragraph{Low-$Z$ regime and analytic treatment.}

For electronic $1s$ states with $Z<8$, and for $2s$ and $2p$ states with $Z<20$,
the finite nuclear size correction becomes extremely small.
In this regime the residual develops a very shallow zero,
making the numerical root-finding procedure unstable.
Increasing precision led to spurious roots and loss of efficiency.

For these specific cases, the analytic finite-size correction formulas
derived by Shabaev~\cite{ref4} were used directly.
In particular, Eqs.~(17)--(19) of Ref.~\cite{ref4}, together with the
coefficients listed in Table~2 of that reference,
were employed to compute
\begin{equation}
E_{\mathrm{FNS}} = E_{\mathrm{0}} + \Delta E_{\mathrm{FNS}}.
\end{equation}
For all other nuclear charges, the full numerical solution of
Eq.~(\ref{eq:energy_condition}) was performed.

\paragraph{Normalization constants and asymptotics.}

Once the eigenenergy was obtained, the radial wavefunctions were normalized
according to
\begin{equation}
\int_{r_{\min}}^{r_{\max}}
\left(G^2(r)+F^2(r)\right)\,dr = 1.
\end{equation}

The matching condition at $r=R_0$ fixes the ratio of the inner and outer
normalization constants, yielding $N_1$ and $N_2$.

In addition, the large-$r$ asymptotic behavior of the outer solution, as defined in Sec.~\ref{sec:boundary_conditions} [Eq.~(\ref{eq:outer_boundary})], was used to extract the exponential prefactor $B$.

The quantities $E$, $N_1$, $N_2$, and $B$ obtained at this stage were stored and used as initial seeds for the fully numerical iterative solver described in Sec.~\ref{sec:fully_numerical_solver}.

\subsection{Fully Numerical Iterative Solver}
\label{sec:fully_numerical_solver}

Although the semi-analytic procedure described in
Sec.~\ref{sec:semi_analytic_solver}
provides reliable reference values for $(E, N_1, B)$, additional stabilization may be
required in regimes where the matching residual becomes nearly flat.
In addition, to treat different nuclear charge-distribution models within a single,
self-consistent framework based directly on integrated wavefunctions, we employ a fully
numerical iterative solver.
The semi-analytic results are used primarily to provide initial seeds for this procedure.

We therefore implement a fully numerical solver in which the
coupled radial Dirac equations
(Eq.~\eqref{eq:coupled_dif_eq})
are integrated from both radial boundaries and stitched at
\begin{equation}
r_{\mathrm{lim}} = R_0,
\end{equation}
with $R_0$ defined by Eq.~(\ref{eq:r0}).
The physical solution is obtained by adjusting
$(E, N_1, B)$ such that continuity and normalization are
simultaneously satisfied.

All numerical calculations in this stage were performed using
arbitrary-precision floating-point arithmetic via the
\texttt{mpmath} library, with a typical setting
$\texttt{mp.dps}=100$ during the iterative solve.

\subsubsection{Runge--Kutta propagation}
\label{sec:rk_propagation}

The radial Dirac system,
Eq.~\eqref{eq:coupled_dif_eq},
is solved using a classical fourth-order Runge--Kutta (RK4) \cite{Hairer1993ODE1}
scheme on the radial grids defined in
Sec.~\ref{sec:radial_grid}.
The inner solution is propagated from
$r_{\min} \to r_{\mathrm{lim}}$,
while the outer solution is propagated from
$r_{\max} \to r_{\mathrm{lim}}$.

The integration step size follows the predefined radial grid,
which is constructed to resolve the nuclear region and
maintain numerical stability.

\subsubsection{Boundary conditions}
\label{sec:boundary_conditions}

\paragraph{Inner boundary ($r = r_{\min}$).}

The regular solution near the origin follows a power-law
behaviour determined by $\kappa$.
Defining $A = 2N_1$, the initial values at $r_{\min}$ are~\cite{Weis2014,Patoary2018}

\begin{align}
G(r_{\min}) &= A\, r_{\min}^{|\kappa|}, \\
F(r_{\min}) &=
\frac{m_0 - E + V(r_{\min})}{|\kappa| + 1 - \kappa}
\, A\, r_{\min}^{|\kappa|+1},
\qquad (\kappa < 0),
\end{align}

and

\begin{align}
F(r_{\min}) &= A\, r_{\min}^{|\kappa|}, \\
G(r_{\min}) &=
\frac{m_0 + E - V(r_{\min})}{|\kappa| + 1 + \kappa}
\, A\, r_{\min}^{|\kappa|+1},
\qquad (\kappa > 0).
\end{align}

\paragraph{Outer boundary ($r = r_{\max}$).}

For bound states ($E < m_0$), the asymptotic solution
decays exponentially.
Using $\lambda$ defined in Eq.~\eqref{eq:rho_lambda},
we set~\cite{Weis2014}

\begin{equation}
\label{eq:outer_boundary}
\begin{aligned}
G(r_{\max}) &= B\, e^{-\lambda r_{\max}}, \\
F(r_{\max}) &= -\sqrt{\frac{m_0 - E}{m_0 + E}}\; G(r_{\max}).
\end{aligned}
\end{equation}

\subsubsection{Matching and residual definition}
\label{sec:matching_residual}

Let $(G_{\mathrm{in}},F_{\mathrm{in}})$ denote the solution
integrated from $r_{\min}$ and
$(G_{\mathrm{out}},F_{\mathrm{out}})$
the solution integrated from $r_{\max}$.
At the matching radius $r_{\mathrm{lim}}$,
continuity requires
\begin{equation}
G_{\mathrm{in}} = G_{\mathrm{out}},
\qquad
F_{\mathrm{in}} = F_{\mathrm{out}}.
\end{equation}
Global normalization requires
\begin{equation}
N = \int_{r_{\min}}^{r_{\max}}
\left(G^2(r)+F^2(r)\right) dr = 1.
\end{equation}
We define the residual vector
\begin{equation}
\mathbf{R}(E,N_1,B) =
\begin{pmatrix}
\Delta G \\
\Delta F \\
\Delta N
\end{pmatrix},
\label{eq:vector_residual}
\end{equation}
where
\setlength{\jot}{10pt}
\begin{align}
\Delta G &=
\frac{G_{\mathrm{in}} - G_{\mathrm{out}}}
{\tfrac{1}{2}(G_{\mathrm{in}} + G_{\mathrm{out}})},
\\
\Delta F &=
\frac{F_{\mathrm{in}} - F_{\mathrm{out}}}
{\tfrac{1}{2}(F_{\mathrm{in}} + F_{\mathrm{out}})},
\\
\Delta N &=
\frac{1 - N}{\tfrac{1}{2}(1 + N)}.
\end{align}
The normalization integral is evaluated numerically
on the combined inner and outer grids.
Convergence is declared when
\begin{equation}
|\Delta G| < \varepsilon,
\qquad
|\Delta F| < \varepsilon,
\qquad
|\Delta N| < \varepsilon,
\end{equation}
with $\varepsilon = 10^{-30}$.
A maximum of 50 iterations is sufficient in practice.

\subsubsection{Three-parameter Newton iteration}
\label{sec:newton_iteration}

The nonlinear system $\mathbf{R}(E,N_1,B)=\mathbf{0}$ is solved using a three-dimensional Newton method,
with $\mathbf{x}=(E,N_1,B)^{\mathsf{T}}$.
At iteration $k$, the Newton correction
$\delta\mathbf{x}_k=(\delta E,\delta N_1,\delta B)^{\mathsf{T}}$
is obtained from the linear system
\begin{equation}
\mathbf{J}_k\,\delta\mathbf{x}_k=-\mathbf{R}_k,
\label{eq:newton_linear_system}
\end{equation}
where $\mathbf{R}_k=\mathbf{R}(E_k,N_{1,k},B_k)$ and the Jacobian matrix is defined as
\begin{equation}
(\mathbf{J}_k)_{ij}
=
\frac{\partial R_i}{\partial x_j}
\bigg|_{\mathbf{x}=\mathbf{x}_k},
\qquad
\mathbf{x}_k=(E_k,N_{1,k},B_k).
\label{eq:jacobian_def}
\end{equation}
(Equivalently, $\delta\mathbf{x}_k=-\mathbf{J}_k^{-1}\mathbf{R}_k$.)
The parameters are then updated according to
\begin{equation}
\mathbf{x}_{k+1}=\mathbf{x}_k+\delta\mathbf{x}_k.
\end{equation}


After each full Newton update, the residual norm $\|\mathbf{R}\|_{2}$ is evaluated.
If $\|\mathbf{R}\|_{2}$ increases compared to the previous iteration, the full step is rejected and a reduced update
\begin{equation}
\mathbf{x}_{k+1}=\mathbf{x}_k+\alpha\,\delta\mathbf{x}_k,
\qquad 0<\alpha<1,
\end{equation}
is applied (simple line search).
In the rare case of a near-singular Jacobian, the update is regularized by switching to a small descent step.
\subsection{Cascaded solver for improved initialization}
\label{sec:cascade_solver}

In a subset of nuclear charges, the semi-analytic reference values used as initial seeds for the fully numerical Newton iteration are not sufficiently close to the true solution.
In these cases the matching residual becomes shallow and the Newton update may step outside the convergence region, preventing convergence within the imposed maximum number of iterations.
Moreover, the semi-analytic reference values are obtained for a specific nuclear charge model; when switching to a different charge distribution, these seeds may become less accurate for certain $Z$ values.
To stabilize the fully numerical iteration in such regimes, we employ a cascaded strategy based on the smooth dependence of $(E,N_1,B)$ on the nuclear charge $Z$.

Once a solution is obtained for a given charge $Z^\star$ in a numerically stable regime, its parameters provide an accurate initial seed for neighboring charges $Z^\star\pm 1$. In practice, we begin from the closest $Z$ value for which the iterative solver converges reliably and propagate the solution stepwise toward the non-convergent regime. For example, in the electronic $1s$ case the cascade starts at $Z=9$ and proceeds downward ($Z=8,7,\dots$). An analogous procedure is applied for the $2s$ and $2p$ states starting from $Z=20$. For muonic systems in regimes where convergence deteriorates (e.g.\ for $2s$ states above $Z\approx 80$), the cascade is performed starting from the nearest convergent $Z$ and stepping toward the target charge.

This strategy significantly improves convergence over a wide range of $Z$ values and reflects the tightly coupled nature of the matching and normalization problem, for which reliable initialization is essential.
\subsection{Hyperfine-structure evaluation}
\label{sec:hfs_postprocessing}

For each converged $(Z,n,\kappa)$ configuration,
the corresponding continuous and normalized radial
wavefunctions were stored and subsequently used
to evaluate the finite nuclear size (FNS) correction
to the magnetic-dipole hyperfine splitting
introduced in Sec.~\ref{sec:hpf}.

The hyperfine interaction depends on the radial integral
\begin{equation}
I_{fg} =
\int_{r_{\min}}^{r_{\max}}
f_{n\kappa}(r)\,
g_{n\kappa}(r)\,
dr,
\label{eq:hfs_radial_integral}
\end{equation}
where the reduced components are defined in
Eq.~\eqref{eq:reduced_components}.

Substituting the general hyperfine expression
(Eq.~\eqref{eq:hfs_general})
and its analytic point-nucleus limit
(Eq.~\eqref{eq:hfs_point})
into the definition of $\delta$
(Eq.~\eqref{eq:delta_definition}),
one obtains
\begin{equation} \delta = 1 - \frac{ X\, j(j+1)\, I_{fg} }{ m_0^2\, Y\, A(Z\alpha) }, \label{eq:delta_final} \end{equation} where \begin{equation} X = \frac{4\kappa}{4\kappa^2 - 1},\qquad Y = \frac{(Z\alpha)^3}{n^3 (2l+1)}. \label{eq:XY_definition} \end{equation}

\subsection{Uncertainty from nuclear charge radii}
\label{sec:uncertainty_propagation}

For the uniformly charged sphere model,
the effective nuclear radius $R_0$ is determined from
the experimental root-mean-square charge radius
$R_{\mathrm{rms}}$ (Sec.~\ref{sec:charge_distribution}).
The quoted experimental uncertainty $\sigma_R$
therefore propagates into the finite nuclear size
correction $\delta$.

To quantify this effect, we apply first-order
error propagation with respect to $R_{\mathrm{rms}}$.
In differential form,
\begin{equation}
\sigma_\delta
=
\left|
\frac{\partial \delta}{\partial R_{\mathrm{rms}}}
\right|
\sigma_R.
\label{eq:error_propagation_general}
\end{equation}

Since the derivative
$\partial \delta / \partial R_{\mathrm{rms}}$
is not evaluated analytically,
it is approximated numerically by a central finite difference,
\begin{equation}
\frac{\partial \delta}{\partial R_{\mathrm{rms}}}
\approx
\frac{
\delta(R_{\mathrm{rms}}+\sigma_R)
-
\delta(R_{\mathrm{rms}}-\sigma_R)
}{
2\sigma_R
}.
\label{eq:finite_difference_derivative}
\end{equation}
The finite-difference step is chosen equal to the quoted experimental uncertainty.

In practice, three independent calculations are performed
using
\begin{equation}
R_{\mathrm{rms}}^{(0)}, \qquad
R_{\mathrm{rms}}^{(\pm)}
=
R_{\mathrm{rms}}^{(0)} \pm \sigma_R,
\end{equation}
yielding corresponding values
$\delta_0$, $\delta_+$, and $\delta_-$.
This leads to the symmetric estimate
\begin{equation}
\sigma_\delta
=
\frac{|\delta_+ - \delta_-|}{2}.
\label{eq:sigma_delta_final}
\end{equation}

The relative uncertainty is reported as
\begin{equation}
\frac{\sigma_\delta}{\delta_0}.
\label{eq:relative_sigma_delta}
\end{equation}

This procedure provides a consistent first-order estimate
of the nuclear-radius-induced uncertainty in the
finite nuclear size correction to the hyperfine splitting.
\section{Numerical Results}
\label{sec:data}

This section summarizes the numerical results for the finite-nuclear-size (FNS) contribution to the Dirac hyperfine correction factor $\delta$ in hydrogenlike systems.
Calculations are performed over a broad range of nuclear charge numbers $Z$ using experimental root-mean-square (rms) charge radii $R_{\mathrm{rms}}$ as input.

For each nucleus, we list the adopted $R_{\mathrm{rms}}$ value with its quoted experimental uncertainty and report the corresponding $\delta$ values for the $1s$, $2s$, and $2p_{1/2}$ states.
Results obtained with the two-parameter Fermi distribution and the uniformly charged sphere model are shown side by side to quantify nuclear-model dependence. The full dataset is summarized in Table~\ref{tab:FNS_delta}.
Uncertainty propagation from $R_{\mathrm{rms}}$ to $\delta$ is evaluated for the uniformly charged sphere model using the first-order procedure described in Sec.~\ref{sec:uncertainty_propagation}.

All radii are given in femtometers (fm) and $\delta$ is dimensionless.
Uncertainties in $R_{\mathrm{rms}}$ are quoted in parentheses and correspond to one standard deviation.

\FloatBarrier
\begin{table*}[t]
\centering
\small
\setlength{\tabcolsep}{8pt}
\renewcommand{\arraystretch}{1.2}

\begin{adjustbox}{max width=\textwidth}
\begin{tabular}{c c
S[mode=math,table-format=1.6e-1, table-align-exponent=false]
S[mode=math,table-format=1.6e-1]
@{\hspace{30pt}}
S[mode=math,table-format=1.6e-1, table-align-exponent=false]
S[mode=math,table-format=1.6e-1]
@{\hspace{30pt}}
S[mode=math,table-format=1.6e-1, table-align-exponent=false]
S[mode=math,table-format=1.6e-1]}
\toprule
$Z$ &
$R_{\mathrm{rms}}$ (fm) &
\multicolumn{2}{c}{$1s$} &
\multicolumn{2}{c}{$2s$} &
\multicolumn{2}{c}{$2p_{1/2}$} \\
\cmidrule(lr){3-4}\cmidrule(lr){5-6}\cmidrule(lr){7-8}
&
&
$\delta_{\rm Fermi}$ & $\delta_{\rm Uniform}$ &
$\delta_{\rm Fermi}$ & $\delta_{\rm Uniform}$ &
$\delta_{\rm Fermi}$ & $\delta_{\rm Uniform}$ \\
\midrule
1  & 0.84075(64)  & 5.984e-3 & 6.4346(49)e-3 & 5.973e-3 & 6.4232(49)e-3 & {} & {} \\
2  & 1.6755(28)  & 2.578e-2 & 2.6156(45)e-2 & 2.560e-2 & 2.5982(44)e-2 & 3.456e-6 & 3.4825(73)e-6 \\
3  & 2.5890(390)  & 5.922e-2 & 6.175(95)e-2 & 5.834e-2 & 6.087(92)e-2 & 2.148e-5 & 2.144(48)e-5 \\
4  & 2.5190(120)  & 7.712e-2 & 8.059(39)e-2 & 7.569e-2 & 7.916(38)e-2 & 4.918e-5 & 4.917(35)e-5 \\
5  & 2.4060(294)  & 9.199e-2 & 9.66(12)e-2 & 9.002e-2 & 9.46(11)e-2 & 9.014e-5 & 9.04(16)e-5 \\
6  & 2.4702(22)  & 1.141e-1 & 1.1931(11)e-1 & 1.112e-1 & 1.1644(10)e-1 & 1.633e-4 & 1.6341(21)e-4 \\
7  & 2.5582(70)  & 1.385e-1 & 1.4422(39)e-1 & 1.345e-1 & 1.4025(37)e-1 & 2.754e-4 & 2.749(11)e-4 \\
8  & 2.6991(52)  & 1.675e-1 & 1.7350(32)e-1 & 1.620e-1 & 1.6805(31)e-1 & 4.495e-4 & 4.474(13)e-4 \\
10  & 3.0055(21)  & 2.319e-1 & 2.3818(15)e-1 & 2.226e-1 & 2.2894(14)e-1 & 1.053e-3 & 1.0439(12)e-3 \\
11  & 2.9936(21)  & 2.526e-1 & 2.5942(16)e-1 & 2.420e-1 & 2.4886(15)e-1 & 1.399e-3 & 1.3874(15)e-3 \\
12  & 3.0570(16)  & 2.792e-1 & 2.8628(13)e-1 & 2.668e-1 & 2.7394(12)e-1 & 1.888e-3 & 1.8707(16)e-3 \\
13  & 3.0610(31)  & 3.005e-1 & 3.0799(27)e-1 & 2.866e-1 & 2.9420(25)e-1 & 2.415e-3 & 2.3939(38)e-3 \\
15  & 3.1889(19)  & 3.531e-1 & 3.6094(17)e-1 & 3.354e-1 & 3.4343(16)e-1 & 3.994e-3 & 3.9566(38)e-3 \\
16  & 3.2611(18)  & 3.799e-1 & 3.8780(17)e-1 & 3.602e-1 & 3.6835(15)e-1 & 5.042e-3 & 4.9947(44)e-3 \\
17  & 3.3654(191)  & 4.095e-1 & 4.173(18)e-1 & 3.876e-1 & 3.957(16)e-1 & 6.385e-3 & 6.323(58)e-3 \\
18  & 3.4028(19)  & 4.321e-1 & 4.4006(18)e-1 & 4.086e-1 & 4.1681(16)e-1 & 7.735e-3 & 7.6613(69)e-3 \\
19  & 3.4349(19)  & 4.537e-1 & 4.6175(18)e-1 & 4.286e-1 & 4.3697(17)e-1 & 9.256e-3 & 9.1697(82)e-3 \\
20  & 3.4776(19)  & 4.758e-1 & 4.8386(18)e-1 & 4.491e-1 & 4.5753(17)e-1 & 1.103e-2 & 1.09328(96)e-2 \\
26  & 3.7377(16)  & 5.959e-1 & 6.0363(14)e-1 & 5.613e-1 & 5.6963(13)e-1 & 2.729e-2 & 2.7082(18)e-2 \\
30  & 3.9283(15)  & 6.647e-1 & 6.7185(12)e-1 & 6.265e-1 & 6.3445(11)e-1 & 4.504e-2 & 4.4743(26)e-2 \\
36  & 4.1835(21)  & 7.471e-1 & 7.5321(14)e-1 & 7.065e-1 & 7.1361(13)e-1 & 8.362e-2 & 8.3232(61)e-2 \\
40  & 4.2694(10)  & 7.863e-1 & 7.91886(59)e-1 & 7.459e-1 & 7.52585(55)e-1 & 1.153e-1 & 1.14929(37)e-1 \\
47  & 4.5454(31)  & 8.472e-1 & 8.5152(14)e-1 & 8.089e-1 & 8.1433(13)e-1 & 1.916e-1 & 1.9137(16)e-1 \\
50  & 4.6519(21)  & 8.674e-1 & 8.71306(81)e-1 & 8.307e-1 & 8.35639(79)e-1 & 2.299e-1 & 2.2971(12)e-1 \\
54  & 4.7964(47)  & 8.906e-1 & 8.9390(15)e-1 & 8.563e-1 & 8.6062(15)e-1 & 2.856e-1 & 2.8563(29)e-1 \\
55  & 4.8041(46)  & 8.948e-1 & 8.9805(14)e-1 & 8.612e-1 & 8.6543(14)e-1 & 2.984e-1 & 2.9852(29)e-1 \\
60  & 4.9110(26)  & 9.157e-1 & 9.18364(66)e-1 & 8.855e-1 & 8.89156(66)e-1 & 3.697e-1 & 3.7000(18)e-1 \\
62  & 5.0819(60)  & 9.260e-1 & 9.2830(13)e-1 & 8.973e-1 & 9.0051(13)e-1 & 4.078e-1 & 4.0826(41)e-1 \\
63  & 5.1115(62)  & 9.294e-1 & 9.3165(13)e-1 & 9.015e-1 & 9.0459(13)e-1 & 4.232e-1 & 4.2375(42)e-1 \\
64  & 5.1319(41)  & 9.325e-1 & 9.34675(83)e-1 & 9.053e-1 & 9.08330(84)e-1 & 4.381e-1 & 4.3866(28)e-1 \\
65  & 5.0600(1500)  & 9.336e-1 & 9.358(30)e-1 & 9.071e-1 & 9.101(31)e-1 & 4.467e-1 & 4.47(10)e-1 \\
66  & 5.2074(172)  & 9.390e-1 & 9.4100(31)e-1 & 9.133e-1 & 9.1611(32)e-1 & 4.702e-1 & 4.709(12)e-1 \\
69  & 5.2256(35)  & 9.462e-1 & 9.47940(57)e-1 & 9.226e-1 & 9.25189(58)e-1 & 5.119e-1 & 5.1276(23)e-1 \\
70  & 5.2995(58)  & 9.495e-1 & 9.51141(88)e-1 & 9.267e-1 & 9.29118(90)e-1 & 5.302e-1 & 5.3110(38)e-1 \\
75  & 5.3698(173)  & 9.598e-1 & 9.6114(21)e-1 & 9.406e-1 & 9.4259(22)e-1 & 6.007e-1 & 6.017(11)e-1 \\
79  & 5.4371(38)  & 9.668e-1 & 9.67993(39)e-1 & 9.504e-1 & 9.52105(40)e-1 & 6.555e-1 & 6.5664(21)e-1 \\
81  & 5.4759(26)  & 9.700e-1 & 9.71094(24)e-1 & 9.549e-1 & 9.56486(24)e-1 & 6.822e-1 & 6.8336(14)e-1 \\
82  & 5.5012(13)  & 9.716e-1 & 9.72607(11)e-1 & 9.572e-1 & 9.58626(12)e-1 & 6.955e-1 & 6.96724(68)e-1 \\
83  & 5.5211(26)  & 9.731e-1 & 9.74014(21)e-1 & 9.593e-1 & 9.60649(22)e-1 & 7.084e-1 & 7.0959(13)e-1 \\
86  & 5.5915(176)  & 9.772e-1 & 9.7799(12)e-1 & 9.652e-1 & 9.6641(13)e-1 & 7.462e-1 & 7.4735(80)e-1 \\
90  & 5.7848(124)  & 9.825e-1 & 9.83127(65)e-1 & 9.729e-1 & 9.73775(67)e-1 & 7.969e-1 & 7.9799(47)e-1 \\
92  & 5.8571(33)  & 9.847e-1 & 9.85213(15)e-1 & 9.761e-1 & 9.76895(16)e-1 & 8.193e-1 & 8.2034(11)e-1 \\
96  & 5.8429(181)  & 9.878e-1 & 9.88277(67)e-1 & 9.812e-1 & 9.81859(67)e-1 & 8.558e-1 & 8.5676(52)e-1 \\
\bottomrule
\end{tabular}
\end{adjustbox}

\caption{Finite nuclear size correction parameter $\delta$ for Fermi and uniform charge distributions in hydrogenlike systems.}
\label{tab:FNS_delta}
\end{table*}

\section{Analysis}
\label{sec:analysis}

\subsection{$Z$-dependence of the FNS correction}

The numerical values of the finite nuclear size (FNS)
correction parameter $\delta$ are summarized in
Table~\ref{tab:FNS_delta}.
For all states considered, $\delta$ increases
monotonically with the nuclear charge number $Z$
for both the two-parameter Fermi distribution
and the uniformly charged sphere model.
This trend reflects the increasing overlap between
the bound-state Dirac wavefunction and the extended
nuclear charge distribution as $Z$ grows.
In muonic systems, the wavefunction becomes increasingly
concentrated in the nuclear region as $Z$ increases,
thereby enhancing the sensitivity of the hyperfine
interaction to finite-size effects.

\begin{figure}[t]
\centering
\includegraphics[width=0.9\linewidth]{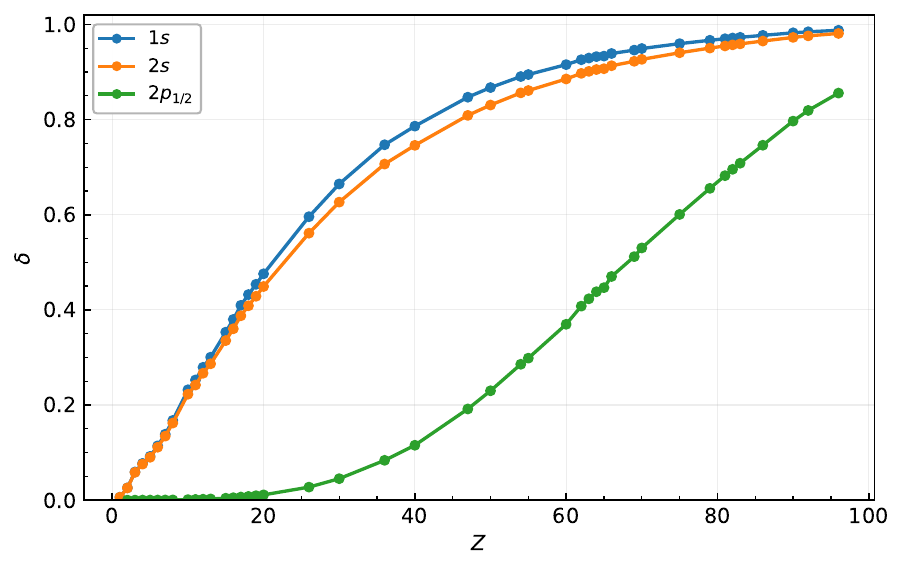}
\caption{Finite nuclear size correction parameter $\delta$
as a function of the nuclear charge number $Z$
for muonic $1s$, $2s$, and $2p_{1/2}$ states
(two-parameter Fermi model). The rapid rise for $s$ states reflects their strong nuclear overlap, while the slower growth of the $2p_{1/2}$ state highlights its relativistic origin.}
\label{fig:delta_vs_Z_fermi}
\end{figure}

The $Z$-dependence obtained with the two-parameter Fermi model
is shown in Fig.~\ref{fig:delta_vs_Z_fermi}.
Since both nuclear models exhibit similar global trends
on this scale, the Fermi results are displayed as representative.

For the $1s$ and $2s$ states, $\delta$ increases rapidly
at low and intermediate $Z$, while the growth rate
progressively decreases for heavy nuclei,
indicating a tendency toward saturation.
In contrast, the $2p_{1/2}$ state exhibits a smaller
absolute magnitude at low $Z$, but displays a
distinct curvature at larger $Z$,
resulting in a qualitatively different $Z$-dependence.

\subsection{State hierarchy}

Across the entire nuclear charge range,
$\delta(1s)$ is consistently slightly larger than $\delta(2s)$,
while both remain of comparable magnitude.
The $2p_{1/2}$ correction is strongly suppressed at low and
intermediate $Z$, consistent with its reduced nuclear overlap,
but becomes increasingly significant for heavy nuclei.
This distinct behaviour highlights the interplay between
relativistic effects and finite-size contributions in
$p_{1/2}$ states.

\subsection{Nuclear-model dependence}

\begin{figure}[t]
\centering
\includegraphics[width=0.9\linewidth]{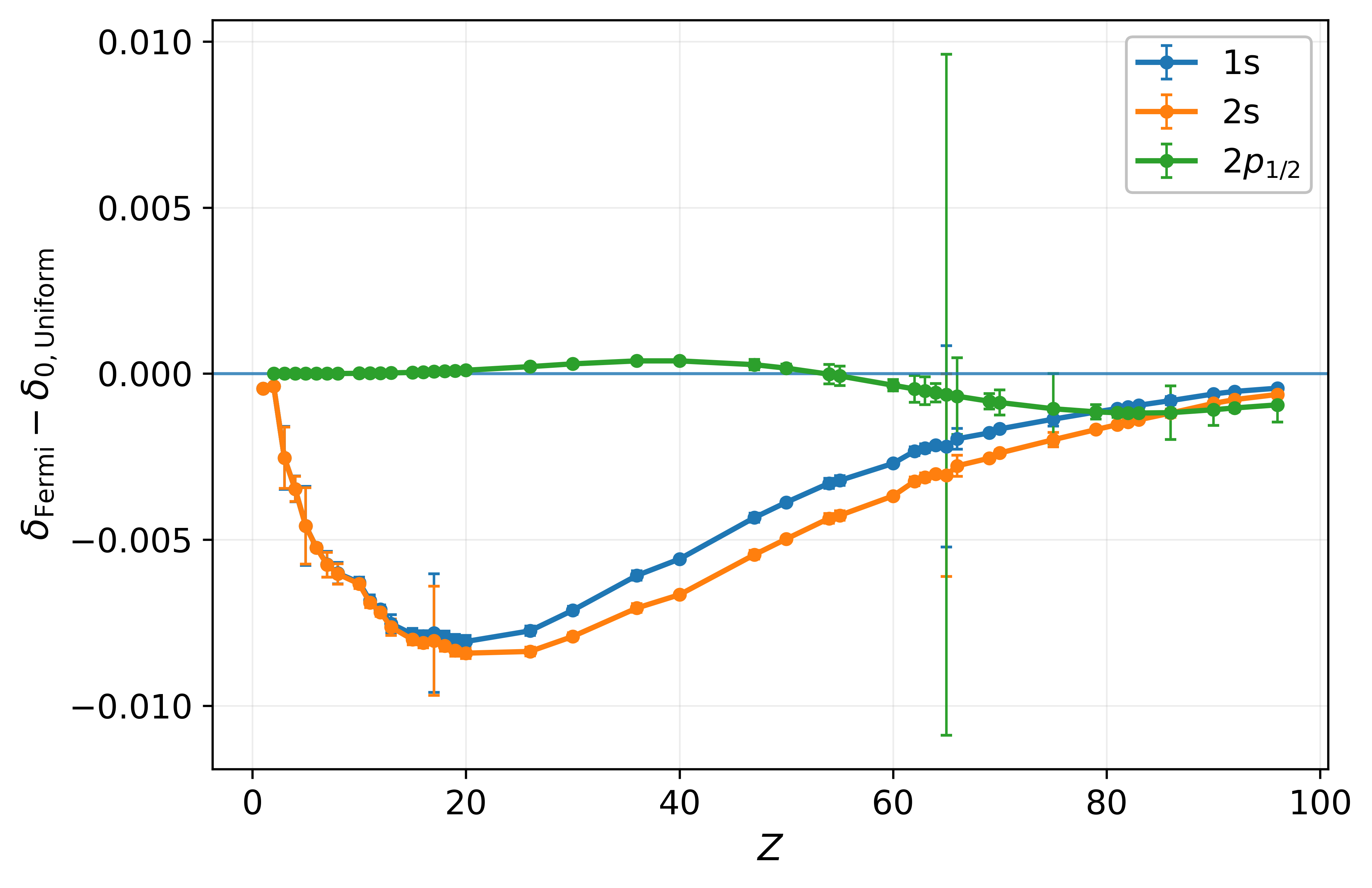}
\caption{Difference between the two-parameter Fermi and uniformly
charged sphere models for the FNS correction parameter $\delta$
in muonic $1s$, $2s$, and $2p_{1/2}$ states.
Error bars represent the propagated rms-radius uncertainty
in the uniform-sphere calculation.}
\label{fig:model_difference}
\end{figure}

To quantify the sensitivity to the assumed nuclear charge
distribution, we define

\begin{equation}
\Delta_{\mathrm{model}}(Z)=\delta_{\mathrm{Fermi}}(Z)-\delta_{0,\mathrm{Uniform}}(Z),
\end{equation}
Here $\delta_{0,\mathrm{Uniform}}$ denotes the central value of the uniform-sphere result evaluated at the adopted rms radius $R_{\mathrm{rms}}^{(0)}$.
We show $\Delta_{\mathrm{model}}(Z)$ in Fig.~\ref{fig:model_difference}.

Error bars indicate the uncertainty propagated from the
experimental root-mean-square (rms) charge radius within the
uniform-sphere model.

For the $1s$ and $2s$ states,
$\delta_{0,\text{Uniform}}$ is systematically larger than
$\delta_{\text{Fermi}}$ over the entire range of $Z$,
corresponding to a negative $\Delta_{\text{model}}$.
The behaviour of the $2p_{1/2}$ state differs qualitatively:
both the magnitude and even the sign of
$\Delta_{\text{model}}$ vary with increasing $Z$,
indicating a stronger sensitivity of the $p_{1/2}$ Dirac
structure to the detailed nuclear surface profile.

For most nuclei, the magnitude of
$\Delta_{\text{model}}$ exceeds the uncertainty propagated
from the experimental rms radius.
Exceptions occur for specific cases,
namely $Z=65$ in the $1s$ state and
$Z=6,\,54,\,55,\,65,\,66,$ and $75$
in the $2p_{1/2}$ state,
where the propagated radius uncertainty becomes
comparable to or larger than the model difference.
A complete numerical listing is provided in
Appendix~\ref{tab:app_model_difference_muon}.

These results demonstrate that, in general,
the nuclear charge-distribution model constitutes
a dominant systematic effect in determining
the FNS contribution to hyperfine splitting.

\subsection{Relative uncertainty analysis}

\begin{figure}[t]
\centering
\includegraphics[width=0.9\linewidth]{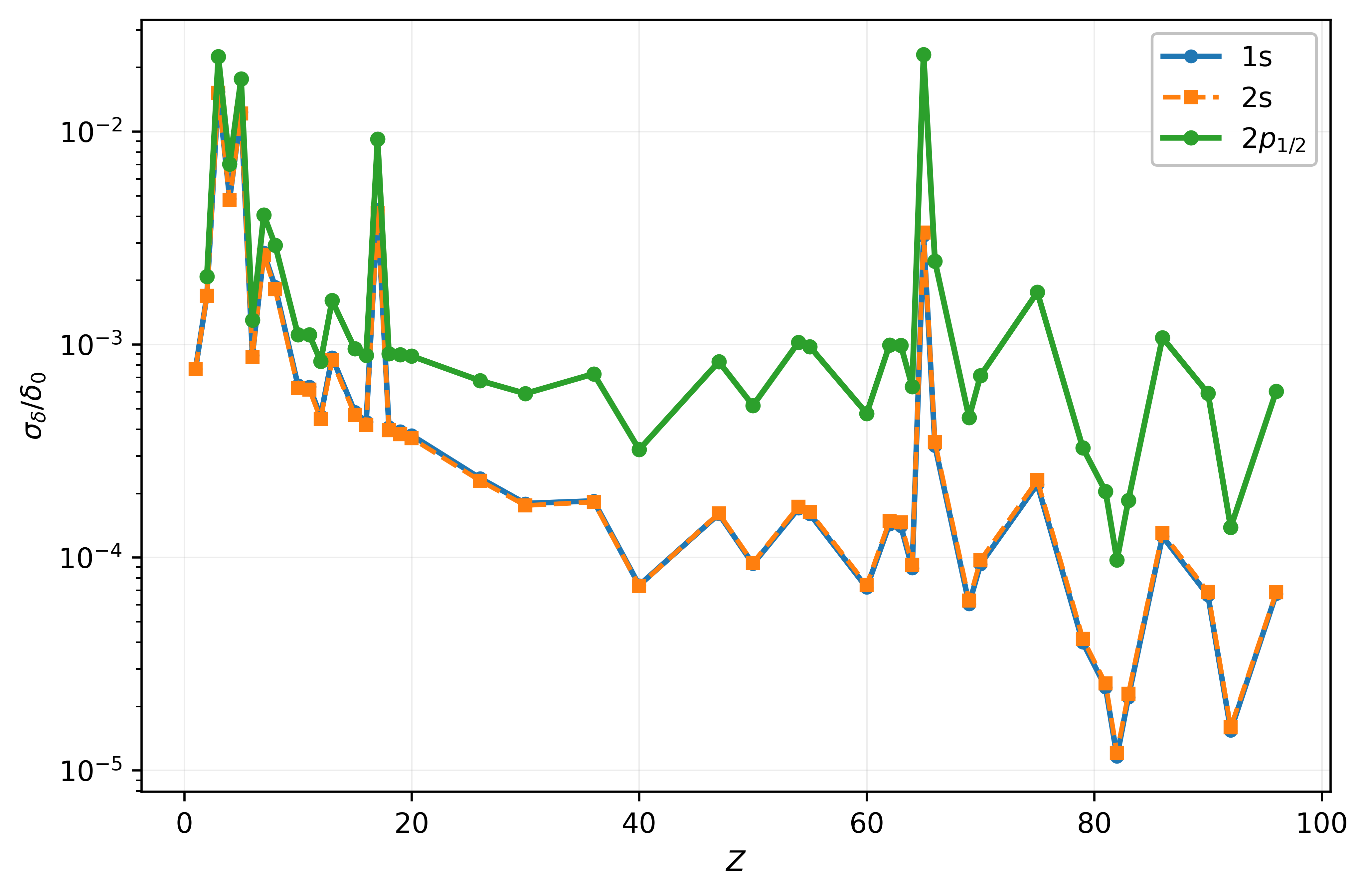}
\caption{Relative uncertainty $\sigma_\delta/\delta_0$
for the uniformly charged sphere model
in muonic $1s$, $2s$, and $2p_{1/2}$ states.
A logarithmic vertical scale is used to display
variations spanning several orders of magnitude.}
\label{fig:relative_uncertainty}
\end{figure}

In addition to absolute uncertainties,
the relative uncertainty
\[
\frac{\sigma_\delta}{\delta_0}
\]
was evaluated for the uniformly charged sphere model.
The results are shown in Fig.~\ref{fig:relative_uncertainty}.
A logarithmic scale is used to accommodate the wide dynamic range across $Z$
and to resolve the near-overlap of the $1s$ and $2s$ curves.

The $1s$ and $2s$ states exhibit nearly identical relative uncertainties over most of the nuclear charge range.
In contrast, the $2p_{1/2}$ state shows a systematically larger fractional uncertainty.
Beyond the smaller absolute magnitude of $\delta$, this behavior is consistent with the fact that the finite-size sensitivity of $p_{1/2}$ states is largely relativistic in origin, arising from the Dirac coupling between the large and small components.
Consequently, variations of the nuclear size parameter can translate into a comparatively larger fractional change in the hyperfine radial integral than for the $s$ states.

A detailed numerical listing of the relative uncertainties for all nuclei considered is provided in Appendix~\ref{tab:app_relative_uncertainty_muon}.

\section{Conclusion}
\label{sec:conclusion}

We have presented a systematic relativistic Dirac analysis of finite nuclear size (FNS) effects on the magnetic-dipole hyperfine splitting in muonic hydrogenlike ions for the $1s$, $2s$, and $2p_{1/2}$ states.
Using high-precision numerical solutions of the radial Dirac equation with two charge-distribution models (two-parameter Fermi and uniformly charged sphere), we quantified the FNS correction factor $\delta$ over a broad range of nuclear charge numbers $Z$ and assessed the impact of nuclear-radius uncertainties within the uniform-sphere model.

A fundamental outcome of this study is that the assumed nuclear charge-distribution profile constitutes a major systematic contribution to $\delta$.
For $s$ states the uniform-sphere model yields consistently larger corrections than the Fermi distribution, whereas for $2p_{1/2}$ the model dependence is more complex and can change sign with $Z$.
For most nuclei, this model difference exceeds the uncertainty propagated from the experimental rms charge radius, indicating that an effective-radius description alone is generally insufficient for precision hyperfine studies~\cite{Valuev2020}.

In addition, the FNS correction shows a common overall increase with $Z$, while its detailed behavior remains state specific.
Although $\delta$ increases smoothly with $Z$ for all states, the detailed behavior differs across the spectrum:
the $1s$ and $2s$ corrections remain close in magnitude, while the $2p_{1/2}$ contribution shows a markedly different curvature and a stronger relative growth toward heavy nuclei.
This highlights the distinct relativistic sensitivity of the $p_{1/2}$ Dirac structure to finite-size effects.

Finally, the radius-induced fractional uncertainty is nearly identical for the $1s$ and $2s$ states but systematically larger for $2p_{1/2}$.
This reflects the fact that even comparatively small absolute FNS contributions can produce sizeable relative variations in states with enhanced sensitivity to the nuclear region.

The numerical framework developed in this work provides a reproducible basis for extending the present analysis to additional states and to further nuclear-structure effects relevant for spectroscopy, including corrections associated with the nuclear magnetization distribution.
\section*{Acknowledgements}

Doğa Yaşar gratefully acknowledges the Max Planck Institute for Nuclear Physics (MPIK) for its hospitality and for providing an excellent research environment. She also thanks Prof.\ Christoph H.\ Keitel for the opportunity to join the Theoretical Physics division. Supported by the Deutsche Forschungsgemeinschaft (DFG, German Research Foundation) Project-ID 273811115—SFB 1225.

\nocite{*}
\bibliographystyle{apsrev4-2}

\onecolumngrid
\clearpage 
\section*{Appendix A: Supplementary tables}
\label{app:supp_tables}

\begin{table}[h!]
\centering
\caption{Model-difference diagnostic for muonic states. We list the difference
$\Delta_{\mathrm{model}}=\delta_{\mathrm{Fermi}}-\delta_{\mathrm{Uniform}}$
together with the propagated uncertainty $\sigma_\delta$ obtained from the
uniform-sphere radius uncertainty propagation. Missing entries indicate that
a value is not available for that state.}
\label{tab:app_model_difference_muon}

\setlength{\tabcolsep}{0.1pt}
\renewcommand{\arraystretch}{1.08}
\small

\sisetup{
  detect-all,
  mode = math,
  scientific-notation = true,
  exponent-base = 10,
  exponent-product = \times,
  tight-spacing = true,
  table-number-alignment = center,
  table-format = -1.6e-2
}

\begin{tabular}{
  c
  S[scientific-notation=true] S[scientific-notation=true]
  S[scientific-notation=true] S[scientific-notation=true]
  S[scientific-notation=true] S[scientific-notation=true]
}
\toprule
& \multicolumn{2}{c}{$1s$} & \multicolumn{2}{c}{$2s$} & \multicolumn{2}{c}{$2p_{1/2}$} \\
\cmidrule(lr){2-3}\cmidrule(lr){4-5}\cmidrule(lr){6-7}
$Z$ &
{$\Delta_{\mathrm{model}}$} & {$\sigma_\delta^{\mathrm{(uniform)}}$} &
{$\Delta_{\mathrm{model}}$} & {$\sigma_\delta^{\mathrm{(uniform)}}$} &
{$\Delta_{\mathrm{model}}$} & {$\sigma_\delta^{\mathrm{(uniform)}}$} \\
\midrule
  1  & -4.507817e-04 & 4.940405e-06 & -4.507293e-04 & 4.923093e-06 & \multicolumn{1}{c}{---} & \multicolumn{1}{c}{---} \\
  2  & -3.794211e-04 & 4.454627e-05 & -3.794573e-04 & 4.397360e-05 & -2.647278e-08 & 7.267869e-09 \\
  3  & -2.535455e-03 & 9.487422e-04 & -2.535976e-03 & 9.235941e-04 &  4.301321e-08 & 4.824422e-07 \\
  4  & -3.470887e-03 & 3.897597e-04 & -3.472275e-03 & 3.770855e-04 &  1.654956e-08 & 3.453343e-07 \\
  5  & -4.584804e-03 & 1.191247e-03 & -4.587221e-03 & 1.147053e-03 & -2.870390e-07 & 1.594767e-06 \\
  6  & -5.236654e-03 & 1.061651e-04 & -5.242713e-03 & 1.015957e-04 & -1.438467e-07 & 2.125031e-07 \\
  7  & -5.743598e-03 & 3.889343e-04 & -5.755420e-03 & 3.699361e-04 &  4.562233e-07 & 1.115262e-06 \\
  8  & -6.012105e-03 & 3.232943e-04 & -6.032571e-03 & 3.055425e-04 &  2.132885e-06 & 1.308378e-06 \\
 10  & -6.285787e-03 & 1.530487e-04 & -6.332704e-03 & 1.429644e-04 &  9.324646e-06 & 1.159590e-06 \\
 11  & -6.830484e-03 & 1.641673e-04 & -6.893104e-03 & 1.528833e-04 &  1.196738e-05 & 1.539446e-06 \\
 12  & -7.101536e-03 & 1.317231e-04 & -7.184256e-03 & 1.222343e-04 &  1.692850e-05 & 1.558443e-06 \\
 13  & -7.531354e-03 & 2.681434e-04 & -7.636168e-03 & 2.482147e-04 &  2.120240e-05 & 3.845846e-06 \\
 15  & -7.846437e-03 & 1.742560e-04 & -8.007956e-03 & 1.604841e-04 &  3.702716e-05 & 3.775676e-06 \\
 16  & -7.908677e-03 & 1.678424e-04 & -8.103559e-03 & 1.542655e-04 &  4.762935e-05 & 4.436936e-06 \\
 17  & -7.813270e-03 & 1.787437e-03 & -8.045184e-03 & 1.639843e-03 &  6.184301e-05 & 5.822882e-05 \\
 18  & -7.927855e-03 & 1.796899e-04 & -8.197481e-03 & 1.646886e-04 &  7.379945e-05 & 6.932310e-06 \\
 19  & -8.032054e-03 & 1.809710e-04 & -8.341419e-03 & 1.657512e-04 &  8.653332e-05 & 8.199282e-06 \\
 20  & -8.066407e-03 & 1.810786e-04 & -8.417673e-03 & 1.657768e-04 &  1.012517e-04 & 9.639230e-06 \\
 26  & -7.745652e-03 & 1.422825e-04 & -8.365787e-03 & 1.306094e-04 &  2.110174e-04 & 1.830271e-05 \\
 30  & -7.127259e-03 & 1.203732e-04 & -7.914880e-03 & 1.111705e-04 &  2.942267e-04 & 2.629646e-05 \\
 36  & -6.079303e-03 & 1.387324e-04 & -7.059395e-03 & 1.298382e-04 &  3.836070e-04 & 6.054366e-05 \\
 40  & -5.582993e-03 & 5.858154e-05 & -6.657053e-03 & 5.532499e-05 &  3.840674e-04 & 3.686190e-05 \\
 47  & -4.337891e-03 & 1.359990e-04 & -5.454643e-03 & 1.310393e-04 &  2.699733e-04 & 1.587008e-04 \\
 50  & -3.875314e-03 & 8.102834e-05 & -4.978618e-03 & 7.874961e-05 &  1.650142e-04 & 1.185879e-04 \\
 54  & -3.300851e-03 & 1.516006e-04 & -4.359466e-03 & 1.490473e-04 & -1.763013e-05 & 2.921595e-04 \\
 55  & -3.217701e-03 & 1.434831e-04 & -4.269629e-03 & 1.413255e-04 & -6.981883e-05 & 2.913554e-04 \\
 60  & -2.704190e-03 & 6.619822e-05 & -3.683395e-03 & 6.591093e-05 & -3.497789e-04 & 1.754513e-04 \\
 62  & -2.337231e-03 & 1.326856e-04 & -3.243435e-03 & 1.333345e-04 & -4.613444e-04 & 4.055828e-04 \\
 63  & -2.241023e-03 & 1.309051e-04 & -3.125816e-03 & 1.318478e-04 & -5.176888e-04 & 4.198099e-04 \\
 64  & -2.158824e-03 & 8.295482e-05 & -3.024068e-03 & 8.370594e-05 & -5.733152e-04 & 2.778201e-04 \\
 65  & -2.190424e-03 & 3.032859e-03 & -3.060523e-03 & 3.052937e-03 & -6.330343e-04 & 1.025349e-02 \\
 66  & -1.961667e-03 & 3.143220e-04 & -2.776923e-03 & 3.187559e-04 & -6.791280e-04 & 1.156707e-03 \\
 69  & -1.785689e-03 & 5.710293e-05 & -2.546979e-03 & 5.805774e-05 & -8.308799e-04 & 2.324038e-04 \\
 70  & -1.663816e-03 & 8.831642e-05 & -2.389809e-03 & 9.014121e-05 & -8.705933e-04 & 3.787538e-04 \\
 75  & -1.371713e-03 & 2.118156e-04 & -1.990805e-03 & 2.169727e-04 & -1.058289e-03 & 1.058178e-03 \\
 79  & -1.154995e-03 & 3.851413e-05 & -1.684774e-03 & 3.950128e-05 & -1.151797e-03 & 2.142835e-04 \\
 81  & -1.052193e-03 & 2.382951e-05 & -1.537015e-03 & 2.444269e-05 & -1.177116e-03 & 1.393321e-04 \\
 82  & -9.997227e-04 & 1.128778e-05 & -1.461440e-03 & 1.158015e-05 & -1.183126e-03 & 6.766656e-05 \\
 83  & -9.519862e-04 & 2.142383e-05 & -1.391787e-03 & 2.197256e-05 & -1.186994e-03 & 1.313808e-04 \\
 86  & -8.125624e-04 & 1.227193e-04 & -1.187165e-03 & 1.257161e-04 & -1.175361e-03 & 8.035480e-04 \\
 90  & -6.130491e-04 & 6.525625e-05 & -8.967741e-04 & 6.699755e-05 & -1.087432e-03 & 4.704381e-04 \\
 92  & -5.358523e-04 & 1.514649e-05 & -7.814383e-04 & 1.552960e-05 & -1.033545e-03 & 1.133473e-04 \\
 96  & -4.386070e-04 & 6.660254e-05 & -6.269955e-04 & 6.723496e-05 & -9.401830e-04 & 5.167722e-04 \\
\bottomrule
\end{tabular}
\end{table}

\begin{table}[t]
\centering
\caption{Relative uncertainty in the finite nuclear size correction,
reported as $\sigma_\delta/\delta_0$ for the uniform-sphere uncertainty
propagation.}
\label{tab:app_relative_uncertainty_muon}

\setlength{\tabcolsep}{6pt}
\renewcommand{\arraystretch}{1.08}
\small

\sisetup{
  detect-all,
  mode = math,
  scientific-notation = true,
  exponent-base = 10,
  exponent-product = \times,
  tight-spacing = true
}

\begin{tabular}{c S[scientific-notation=true] S[scientific-notation=true] S[scientific-notation=true]}
\toprule
$Z$ & {$\sigma_\delta/\delta_0\ (1s)$} & {$\sigma_\delta/\delta_0\ (2s)$} & {$\sigma_\delta/\delta_0\ (2p_{1/2})$} \\
\midrule
  1  & 7.67782090e-04 & 7.66450800e-04 & \multicolumn{1}{c}{---} \\
  2  & 1.70307504e-03 & 1.69249291e-03 & 2.08698525e-03 \\
  3  & 1.53633469e-02 & 1.51726896e-02 & 2.25013471e-02 \\
  4  & 4.83647185e-03 & 4.76370451e-03 & 7.02397701e-03 \\
  5  & 1.23347684e-02 & 1.21247859e-02 & 1.76359571e-02 \\
  6  & 8.89855617e-04 & 8.72491271e-04 & 1.30043711e-03 \\
  7  & 2.69674255e-03 & 2.63775287e-03 & 4.05681409e-03 \\
  8  & 1.86333963e-03 & 1.81816871e-03 & 2.92452408e-03 \\
 10  & 6.42586098e-04 & 6.24458141e-04 & 1.11086656e-03 \\
 11  & 6.32813216e-04 & 6.14339191e-04 & 1.10959954e-03 \\
 12  & 4.60123849e-04 & 4.46210939e-04 & 8.33065551e-04 \\
 13  & 8.70620497e-04 & 8.43688762e-04 & 1.60655030e-03 \\
 15  & 4.82777909e-04 & 4.67291525e-04 & 9.54282230e-04 \\
 16  & 4.32808768e-04 & 4.18800650e-04 & 8.88327501e-04 \\
 17  & 4.28334089e-03 & 4.14439963e-03 & 9.20942524e-03 \\
 18  & 4.08328385e-04 & 3.95114990e-04 & 9.04849818e-04 \\
 19  & 3.91921633e-04 & 3.79317113e-04 & 8.94166702e-04 \\
 20  & 3.74234935e-04 & 3.62328747e-04 & 8.81683053e-04 \\
 26  & 2.35712778e-04 & 2.29289551e-04 & 6.75827489e-04 \\
 30  & 1.79167126e-04 & 1.75222911e-04 & 5.87717273e-04 \\
 36  & 1.84187369e-04 & 1.81946505e-04 & 7.27407211e-04 \\
 40  & 7.39772099e-05 & 7.35132863e-05 & 3.20737269e-04 \\
 47  & 1.59713769e-04 & 1.60916687e-04 & 8.29279101e-04 \\
 50  & 9.29963978e-05 & 9.42387404e-05 & 5.16260637e-04 \\
 54  & 1.69593755e-04 & 1.73185602e-04 & 1.02285363e-03 \\
 55  & 1.59771008e-04 & 1.63300751e-04 & 9.76000598e-04 \\
 60  & 7.20827636e-05 & 7.41275156e-05 & 4.74191632e-04 \\
 62  & 1.42934009e-04 & 1.48064775e-04 & 9.93440631e-04 \\
 63  & 1.40508542e-04 & 1.45754695e-04 & 9.90701419e-04 \\
 64  & 8.87525973e-05 & 9.21536408e-05 & 6.33332279e-04 \\
 65  & 3.24109827e-03 & 3.35446553e-03 & 2.29216644e-02 \\
 66  & 3.34028972e-04 & 3.47945488e-04 & 2.45638882e-03 \\
 69  & 6.02389633e-05 & 6.27522708e-05 & 4.53240395e-04 \\
 70  & 9.28531614e-05 & 9.70180639e-05 & 7.13150519e-04 \\
 75  & 2.20380674e-04 & 2.30188335e-04 & 1.75854236e-03 \\
 79  & 3.97876077e-05 & 4.14883456e-05 & 3.26332941e-04 \\
 81  & 2.45388249e-05 & 2.55546687e-05 & 2.03893813e-04 \\
 82  & 1.16056899e-05 & 1.20799454e-05 & 9.71210105e-05 \\
 83  & 2.19953961e-05 & 2.28726297e-05 & 1.85149629e-04 \\
 86  & 1.25480798e-04 & 1.30085528e-04 & 1.07519864e-03 \\
 90  & 6.63761943e-05 & 6.88018675e-05 & 5.89530092e-04 \\
 92  & 1.53738216e-05 & 1.58968998e-05 & 1.38171508e-04 \\
 96  & 6.73925588e-05 & 6.84772121e-05 & 6.03167894e-04 \\
\bottomrule
\end{tabular}
\end{table}

\end{document}